# Experience in engineering of scientific software: The case of an optimization software for oil pipelines

Vahid Garousi
Queen's University Belfast
Northern Ireland, UK
v.garousi@qub.ac.uk

Ehsan Abbasi
AltaLink, Calgary, Canada
eabbasi@ucalgary.ca

Bedir Tekinerdogan
Information Technology Group
Wageningen University, Netherlands
bedir.tekinerdogan@wur.nl

**Abstract**: Development of scientific and engineering software is usually different and could be more challenging than the development of conventional enterprise software. The authors were involved in a technology-transfer project between academia and industry which focused on engineering, development and testing of a software for optimization of pumping energy costs for oil pipelines. Experts with different skillsets (mechanical, power and software engineers) were involved. Given the complex nature of the software (a sophisticated underlying optimization model) and having experts from different fields, there were challenges in various software engineering aspects of the software system (e.g., requirements and testing). We report our observations and experience in addressing those challenges during our technology-transfer project, and aim to add to the existing body of experience and evidence in engineering of scientific and engineering software. We believe that our observations, experience and lessons learnt could be useful for other researchers and practitioners in engineering of other scientific and engineering software systems.

## 1 Introduction

Scientific software (or "research software") is a certain type of software, which enable scientific computing, and is broadly defined as "*software used or developed for scientific purposes*" [1]. Since engineering and science are closely-related, engineering software is a very similar concept and refers to any software system used or developed for engineering purposes. In almost all areas of science and engineering, scientists and engineers use and/or develop complex software systems nowadays.

It has been widely reported in the literature that development of scientific and engineering software is usually more challenging than the development of conventional software (such as e-business applications) [2-4]. Various reasons have been discussed for those challenges, e.g., difference of skillset and vocabulary between software engineers and domain scientists (or engineers), complex nature of scientific and engineering software which makes almost all phase of the software development lifecycle (SDLC), e.g., requirements, development and testing [5], more challenging compared to conventional software.

We were involved in an industry-academia technology-transfer project on developing a software for optimization of energy costs of pumps in oil pipelines [6]. The project was funded by a governmental funding agency in Canada [6]. Two companies and one university research team were involved in the project. Given the multi-disciplinary nature of the project, experts from different areas participated in the project. We had mechanical engineers, chemical engineers and software engineers in the team. Given the complex nature of the software (which included a sophisticated state-of-the-art computational and optimization model [7, 8]), there were various challenges in different software engineering (SE) aspects of the software system during the project which raised several questions including: how to ensure to capture all (most of) the requirements early on? How to test the optimization sub-system? How to ensure all the engineers (software, mechanical, civil, etc.) would understand each other properly? The latter issue related to the differences in terminologies and domain expertise, when software engineers "meet" other scientists and engineers [9].

In this experience paper, we report our SE observations, experience, challenges and the adopted approaches during the technology-transfer project. Our approach in this experience paper is to present our observations, experience and lessons learned, relate them to the experience and findings reported in the literature, e.g., [10-13], and then to generalize, when possible, the recommendations for practitioners and researchers. Likewise, we aim to add to the existing body of experience in development of scientific and engineering software.

We have structured the rest of this paper as follows. We review the related work in Section 2. We describe an overview of the domain, project and context in Section 3. Section 4 discusses the research method that we have used for synthesizing and reporting the observations, experience and lessons learnt from our project. In Section 5, we present the software engineering activities and the experience in the SE process. Section 6 presents the summary of the paper and the discussions. Finally, in Section 7, we draw conclusions, and discuss areas of future work.

## 2 BACKGROUND AND RELATED WORK

A large number of studies have been published in the area of development of scientific and engineering software. For example, a systematic literature mapping study [14], published in 2013, reviewed and classified 130 papers published in this area. A Systematic Literature Review (SLR) reviewed 43 studies and extracted the claims about the use of SE practices in computational science and was published in 2015 [15]. A SLR of testing approaches for scientific software was published in 2014 [5], which reviewed a set of 49 papers. Another SLR [16], published in 2011, systematically selected nine papers which have reported the usage of agile practices and their effects in scientific software development and systematically reviewed them.

Since the above literature review studies have already reported compressive reviews, we do not intend to provide again a detailed literature review in this paper. We present two insightful charts in Figure 1, which are taken from the systematic literature mapping presented in [14]. The left chart shows the frequency of SE activities/techniques mentioned in the pool of 130 papers in that mapping study. The right chart shows the frequency of application domains in the papers. As we can see, architecture / design, and the development / coding have been the most widely discussed SE issues, as of 2013 [14]. Testing of scientific software has also been an important issue since it has been discussed in 38 papers.

In terms of application domains, physics (37 papers), math (20 papers) and biology and HPC (High Performance Computing), each discussed in 14 papers, have been the fields in which the most papers in this topic (development of scientific software) have been published. The interested reader can refer to [14], and the other SLRs [5, 15, 16] for details.

In the application domains similar to that of the current paper, several experience reports like this paper have been published in the past, e.g., [10-13]. The paper [10] was an experience paper on developing oil-reservoir simulation software. The paper emphasized that software for oil exploration and production has been a topic for software R&D for decades. The role of such applications cannot be underestimated. For example, the paper stressed the high cost of a defect if reservoir drilling is done in the wrong place as a result of a software defect, and knowing that without software, engineers would not have a clue where to start drilling. The paper was written by two practitioners in Shell, which is one of the largest oil companies in the world. The paper provided Lines of Code (LOC) growth charts together with highlights of the releases and also the bug trends chart of a major reservoir simulator software developed in the company. It was discussed that certain defects could be dangerous, e.g., faulty physics or functionality that can lead to erroneous simulation results. The paper explained the software development practices used in the teams and that those practices aim to follow industry best practices, but like any team, they have their own peculiarities. The paper also reported that the team has numerous types of testing in place, such as unit, regression, integration, and performance testing.

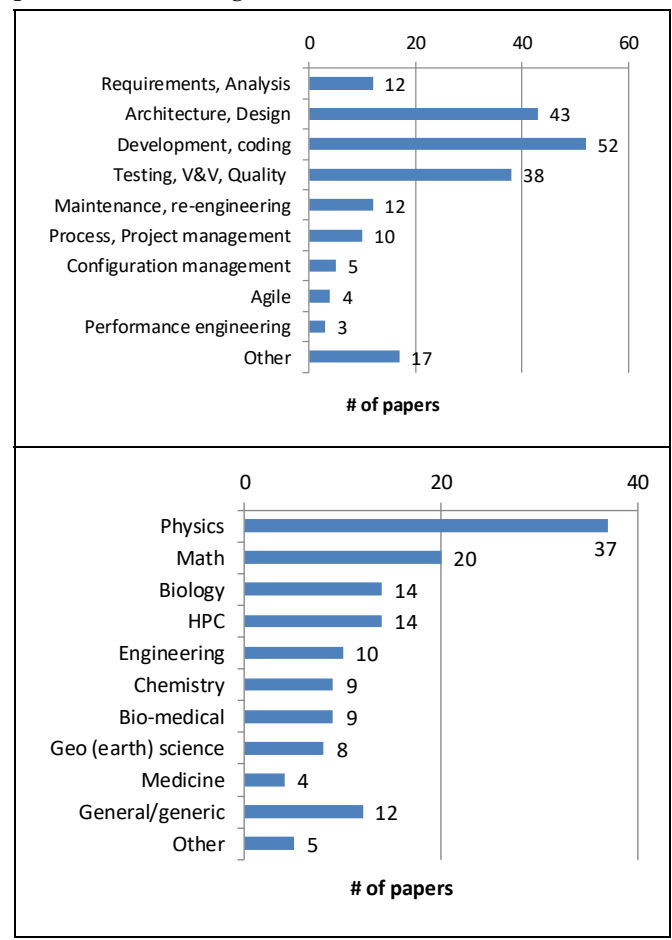

**Figure 1: Two charts from the systematic literature mapping [14]: (Top): Frequency of SE activities/techniques mentioned in the pool of papers; (Bottom): Frequency of application domains in the papers**

Another paper [11] reported the experience and lessons learned in engineering a software for understanding climate change. The authors described an ethnographic study of the culture and practices of climate scientists in a British government agency (the Met Office Hadley Centre for Climate Science and Services). The study examined how the scientists think about software correctness, how they prioritize requirements, and how they develop a shared understanding of their models. The findings showed that climate scientists have developed customized techniques for verification and validation that are tightly integrated into their approach to scientific research. Their software practices share many features of both agile and open-source projects, in that they rely on self-organisation of the teams, extensive use of informal communication channels, and developers who are also users and domain experts.

An invited talk [12] for the American National Center for Supercomputing Applications (NCSA) shared experience and lessons learned from developing scientific software. The software engineering initiative of the German Aerospace Center (DLR) and how their software engineers and scientists overcame the obstacles towards developing sustainable software were reported in [13].

The current paper share similarities, in its focus, with the above experience papers, and contributes to the body of experience in development of scientific and engineering software. Our paper also falls in the area of empirical and evidence-based SE (EBSE) [17] as we are sharing the evidence that we have accumulated during a multi-year software project in the context of an engineering software. In EBSE, researchers draw from their experience in applied SE project and report what works in certain contexts [18] and what does not work.

## 3 AN OVERVIEW OF THE DOMAIN, PROJECT AND CONTEXT

It is important to provide contextual information (factors) in experience papers [18, 19]. Hence, in this section we provide an overview of the domain and the project topic together with the corresponding partner companies and team members.

### 3.1 AN OVERVIEW OF THE DOMAIN AND THE PROJECT TOPIC

The latest data from 2014 [20] gives a total of slightly less than 2,175,000 miles (3,500,000 km) of pipeline in 120 countries of the world. The United States had 65%, Russia had 8%, and Canada had 3%, thus 75% of all pipeline were in these three countries. There are different types of pipelines for transporting different products and liquids, e.g., oil, water. Development, planning and operation of pipelines and pipeline networks are complex tasks and a discipline of its own [21].

Software systems are heavily used in all those aspects, e.g., Supervisory Control and Data Acquisition (SCADA) have long been used during operations for monitoring and control of pipelines [22]. Furthermore, the energy required to operate pump stations in oil pipeline networks can account for enormous energy consumption (either electrical or fossil fuels). Expenses due to electricity usage of pump stations could surpass 50% of operating cost in case of many oil pipeline companies [23]. Furthermore, the global carbon footprint of such a major transportation activity world-wide is in mega scales [23].

To address the above needs, an industry-academia collaborative R&D project was defined and was funded by a governmental funding agency in Canada (details about the consortium and the team is provided in the next subsection). The project title was: "*Engineering intelligent software systems for improving the operational efficiency of oil pipeline networks*", and it ran for five years: 2017-2012.

During the project, the team developed several optimization approaches and a software system for determining optimal pump operation schedule (On/Off schedule and speed of each pump in the network in each time instance of system operation) with multi-tariff electricity supply [8, 24]. Multi-tariff electricity supply in this context means getting different cost values for energy from real contracts with energy providers, in which unit cost of energy would change in different time periods, e.g., it would be less expensive during nighttime (1-6 PM) than day time. Thus, the optimization software has to utilize less expensive times "more" in its output schedule.

The objective of the optimization software was to find the optimal on/off schedule and speed of the operating pumps in the network (for the case of variable speed pumps) in each time instance, such that the sum of energy consumed by all pumps is the least expensive, while all hydraulic, physical and operational constraints of the pipeline are adhered to (e.g., minimum and maximum allowed pressure of each pipeline segment, and hydraulic model). The optimization problem at hand was a complex task as it included the extended-period hydraulic model represented by algebraic equations as well as mixed-integer decision variables [8, 24].

As it is the same with any computational science project, the project had two parallel efforts: (1) carrying out the "domain" research (development of the optimization approaches for the problem at hand); and (2) engineering the software system to utilize the optimization approaches and integrate with the existing systems in the context of industrial partners to provide the optimal pump operation schedule.

The scientific software was developed to serve (support) the domain research [2]. In our case, the domain research involved development and refinement of a computational optimization model, and was the technical "core" on which the scientific software was built up on. As we discuss in Section 3.2, people with the "right" expertise were involved in the domain research activities (mechanical and chemical engineers) and there was active communication along all team members, since we wanted to develop one single software product at the end.

For the domain research, we formulated and solved the optimization problem using different optimization approaches, e.g.: (1) an Mixed-Integer Linear Programming (MILP) [24], and (2) a non-linear model based on Genetic Algorithms [8]. The proposed model was iteratively evaluated and refined on several real-world and hypothetical oil pipeline networks. We also assessed the accuracy and computational performance of the optimization model via a set of experimentations. To keep our focus on SE aspects in this paper, we do not elaborate further on the domain research aspects in this paper, but the interested reader could refer to the papers that we have published on those aspects, e.g., [7, 8, 24].

## 3.2 THE TEAM: SOFTWARE ENGINEERS MEET OTHER SCIENTISTS AND ENGINEERS

The project team was set up with several partners: the principal investigator (PI) (the first author of the paper), two pipeline companies (domain experts serving the role of "problem owners"), one civil-engineering consulting firm (with specialization in optimization), and the PI's research team which was composed of: one masters' student in software engineering, one masters' student and one post-doctoral fellow both with expertise in optimization.

While having experts from different domains (mechanical, power and software engineers) was important for the success of our project, it made collaboration challenging at times, e.g., the were clear differences in terminologies used in the meetings and domain expertise among software engineers and domain scientists / engineers [9] (more details and examples in Section 5).

There have been studies on interaction and collaboration of software engineers with other scientists and engineers [9], which have reported various challenges, such as: upfront articulation of requirements being not trivial; requirements are emergent; and the project documentation alone not being enough to construct a shared understanding between various team members. To ensure a smooth work environment among all the experts in the project, we started with a thorough review of the existing studies in this area, e.g., [9], and recommendations on how to best collaborate, and utilized them during our project.

## 4 RESEARCH METHOD

We used two research methods in our technology-transfer project: action-research [25, 26] and participant-observation [27]. As we discuss next, these methods helped us ensure rigor in our project and this experience report.

Similar to our other past projects, e.g., [28], the project was driven based on principles of technology-transfer, action-research [26] and empirical methods, in which research challenges were dealt with and solved as we met them during the project. We used the guidelines for technology transfer [25] and the action-research research method [29]. The action-research methodology is regarded as "*the most realistic research setting found, because the setting of the study is the same as the setting in which the results will be applied for a given organization, apart from the presence of the researchers*" [30] and its application emphasizes "*more on what practitioners do than what they say they do*" [26]. These characteristics come from a background which is based on the assumption that theory and practice can be closely integrated by learning from the results of intervention that are planned after a thorough diagnosis of the problem context [31].

To systematically design and conduct AR studies in a rigorous manner, a taxonomy was provided by Santos and Travassos in [32]. To put our AR approach in context, we provide the classification of our AR approach using that taxonomy in Table 1.

For each attribute in this taxonomy, we provide the value (or the chosen "mode") in our AR project. Note that this research design is focused on our goal in this paper, which is to extract and synthesize our observations, challenges, experiences and lessons learned in the project, and not the technical project itself (engineering of an optimization software for oil pipelines). Thus, we have phrased the problem (goal) and action field in Table 1 as mentioned.

In terms of "adherence" in Table 1, our chosen mode was "Based", since for our context, it was better to combine the AR methodology with participant-observation and case-study research methods. In terms of "type", our chosen mode was "participatory action research", since given the multi-disciplinary nature of our team, we wanted to emphasize participant collaboration.

For data collection, we aimed at collecting both quantitative and qualitative data. For quantitative data, we gathered metrics such as number of pass/failed test cases during testing, and number of requirements items changed in each iteration of software development (we used the iterative development model). For qualitative data, we documented our observations as diary reflections and kept the history of emails exchanged among team members.

**Table 1 - Classification of our action-research approach based on the action-research taxonomy provided by Santos and Travassos [32].**

| Attribute | | Value |
|---|---|---|
| Problem (goal) | | To extract and synthesize our observations, challenges, experiences and lessons learned in the project (engineering of an optimization software for oil pipelines) |
| Action | | A multi-disciplinary team was formed. We planned and conducted all SE activities (requirements engineering, development and testing) using best practices and empirical evidence in the literature, while documenting our experiences, challenges and lessons learned. We then synthesized retrospectively those experiences, challenges and lessons learned. |
| Adherence | | • Inspired – when the focus is on the researchers learning from a real problem resolution exploring SE research without controlling the study by the AR principles;<br>• Based – when the AR methodology is modified or combined with other empirical methods;<br>• Genuine – when the full essence of action research methodology is present |
| Type | | • Action Research – focusing on change and reflection<br>• Action Science – trying to solve conflicts between "espoused" and applied theories. (Espoused theory is the perspective on what we say, as compared to theory-in-use which are the behind what we do [33].)<br>• Participatory Action Research – emphasizing participant collaboration<br>• Action Learning – for programmed instruction and experiential learning |
| Length | | Five years |
| Data collection | | Quantitative and qualitative were used/ Techniques: Metrics (for quantitative data) and observation (for qualitative data) |
| Action-research control structures | Initialization | • Researcher – field experiment<br>• Practitioner – classic action-research genesis<br>• Collaboration – evolves from existing interaction |
| | Authority | • Practitioner – consultative action warrant<br>• Staged – migration of power to the client<br>• Identity – practitioner and researcher are the same |
| | Formalization | • Formal – specific written contract<br>• Informal – broad, perhaps verbal, agreements<br>• Evolved – informal or formal shift into opposite form |

In terms of action-research control structures, initialization of AR was of "collaboration" type, since it evolved from existing interaction among tam members. "Authority" of AR was of "identity" type, since practitioners and researchers were placed in the same level of authority. In terms of "formalization", the choice was "formal" since there was a written contract for the project itself (funding proposal), in which one of the items in the work packages was to extract and synthesize the SE observations, challenges, experiences and lessons learned in the project.

Our approach for synthesizing and reporting our experience has been to gather the insightful pieces of our experience which we thought will be helpful for other researchers and practitioners in engineering of scientific and engineering software in other contexts. Our experience synthesis and reporting our approach in this paper is similar to the past experience papers that we have published from our SE activities in practice in the past, e.g., [34-36].

Furthermore, the research method that we used for synthesizing and reporting the observations, experience and lessons learnt, in the rest of this paper, is "participant-observation" [27]. This is a method which has been widely used in anthropology, in which the researcher is both an observer and a participant in some activity over time. Participant-observations have the unique strength of describing complex aspects of cognition, social interaction, and culture over time;

The participant-observation method has also been used in SE research, e.g., [37-41]. As an example of a participant-observation study, in 1989, Donald Knuth published the Errors of TeX [41], a diary study documenting and reflecting on the 850 errors in the TeX toolset that he had made over a decade of work. As another example, Ko [37] reported a three-year participant-observation of software evolution in a software startup. The author [37] had more than 9,000 hours of direct experience in a software startup. He analyzed his emails and diary reflections to synthesize a set of nine claims about SE in software startups.

By following the participant-observation method and its guidelines [27], we were able to conduct a rigorous and systematic study in this work. We had kept diary reflections and history of emails during our projects. By systematically analyzing those sources, we synthesize in this paper our experience and to contrast our experience with the many lessons learned in the literature about development of scientific and engineering software [2-4].

## 5 SOFTWARE ENGINEERING ACTIVITIES AND THE EXPERIENCE IN THE PROCESS

We now review the SE process and activities during each of the phases (e.g., requirements and testing), and also synthesize and present the experience in each of the phases.

### 5.1 DEVELOPMENT PROCESS: DOMAIN (SCIENTIFIC) RESEARCH PROCESS + SOFTWARE ENGINEERING PROCESS

According to the literature [16, 42], Agile methods seem to be more popular (suitable), in general, for scientific software development than other software development processes. For example, a literature review of Agile practices in development of scientific software [16] mentioned that: "*The nature of scientific research and the development of scientific software have similarities with processes that follow the agile manifesto: responsiveness to change and collaboration are of the utmost importance*".

In our initial project planning phase, we found various reasons to adopt Agile and iterative development models in our project, e.g., full specification of requirements was not possible in the beginning for the project and it was clear that requirements would be emergent and change during the project. We show in Figure 2 an overview of the software development process, which is inspired from previous works [9, 43] and our own experience in this project. The software product, that we developed, was named *OptimalPipeline*.

The process has two parallel and intertwined sub-processes: domain research (pipeline optimization) and scientific SE, which proceed "hand in hand". Iterations are an inherent aspect of both sub-processes. In principle, refinements and improvements in the domain research drove the need for software maintenance on the scientific software that we were developing, which in turn, serves ("supports") the domain research.

Domain (scientific) research has three activities: (1) Inception of scientific idea, (2) Formulation of initial optimization model, and (3) Refinements, improvements, new features. The SE process consisted of engineering (development and testing) of the subject software, and software maintenance when there were refinements or improvements in the computational approaches. We discuss next our experience in different SE phases, e.g., requirements, development and testing.

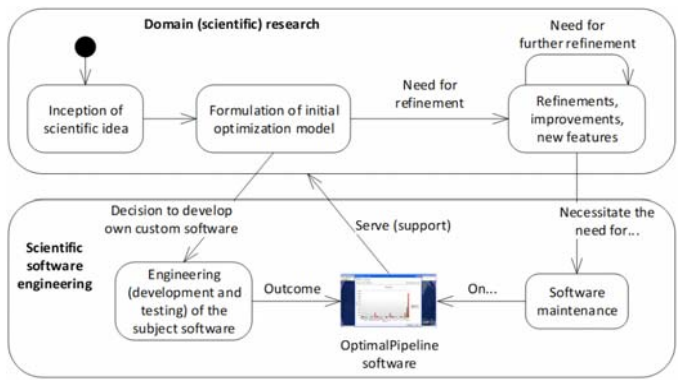

**Figure 2- An overview of the software development process**

**Observation 1:** The development process of most scientific software systems includes two parallel sub-processes: Domain (scientific) research process and software engineering process, which are closely integrated. Careful alignment and synchronization of the two sub-processes is important to ensure success in these projects.

## 5.2 REQUIREMENTS ENGINEERING

Similar to any software project, the first activity was to determine the requirements and scope of the *OptimalPipeline* software. We started with use-case modeling and determined the essential use-cases of the system as shown in Figure 3. During the analysis and design phase, we found that it was important for the OptimalPipeline software to utilize features from four external systems: (1) the map feature from Google Earth, (2) the optimization feature from a popular commercial optimization engine called LINGO (www.lindo.com); (3) the charting feature from Microsoft Excel; and (4) pipeline and pump operational data provided by SCADA systems, which were already in use by the pipeline companies in the project consortium. The links (interfaces) with external systems are depicted in Figure 3.

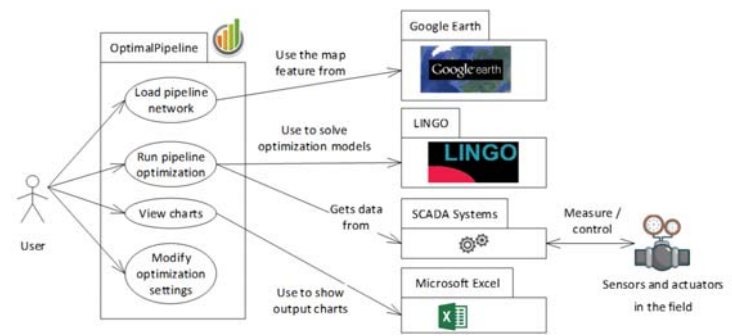

**Figure 3-Use-case diagram of the OptimalPipeline software**

We observed that determining the essential use-cases of the system was relatively easy (as per the use-case diagram shown in Figure 3. However, we observed that most of the requirements complexity was not in figuring out the use-cases themselves, but instead in the domain-specific features and requirements of the system. For example, in the context of our project, we raised and explored the following questions during the requirements engineering phase:

- How should we formulate the optimization problem?
- What types of parameters should be included in the optimization model?
- If the optimization problem is non-linear (which it was), what types of solution mechanisms should we use (e.g., genetic algorithms)?
- Which parameters (characteristics) for pump should be included, e.g., pressure, capacity (flow- rate) and efficiency?

To document the domain-specific features and requirements (such as the above examples), we did not see the need to use a specific requirement engineering language or notation, but we ensured documenting them in the natural language format as precise as possible. After each round of requirements gathering among team members, we ensured peer reviewing them with the domain experts (e.g., mechanical engineers in the partner companies).

Answering and analyzing such questions made all the SDLC phases (and requirements analysis in specific) more interesting and more challenging. Furthermore, extra- (non-) functional requirements of the domain were critical and had to be considered carefully, e.g., each of the oil pipeline segments had minimum and maximum allowed pressure values which were due to physical properties of the pipeline and the terrain in which it was built in. If the pressure of the liquid oil would go above the maximum or below the minimum threshold, catastrophic failures could occur, e.g., leakage and even explosion in the pipeline and pumping systems. Thus, the software had to fully satisfy such non-functional requirements.

These critical types of requirements and possible defect arising from them were quite similar, in nature, to "dangerous" defects as reported in another experience paper [10] (as reviewed in Section 2), e.g., faulty modeling of physics or functionality that can lead to erroneous simulation results.

An important requirements engineering goal for us was to capture all (or most of) the requirements early on. Given the complex nature of the domain and the problem at hand (Section 2.1), this was not really possible, as we could only get more requirements as we were exploring the problem domain (the scientific problem itself). Thus, we were only able to gather the problem's requirements and also software requirements in iteration and thus Agile practices showed to be more applicable in the requirements engineering aspects of this project. We used heuristics from the Agile requirements management literature [16] to help us in cost-effective management of requirements. For example, instead of documenting the requirements in-length, the team (especially non software engineers) preferred "just enough" requirements [44]. We also observed that the practitioner members of the team preferred to "satisfice" (pursue the minimum satisfactory condition or outcome), which has been reported to be quite the case among most practitioner software engineers [45], and also other practitioner in other areas of engineering, e.g., in civil engineering [46].

**Observation 2:** In scientific software projects, domain-specific requirements could often be more complex than software-specific requirements (e.g., use-case descriptions).

As discussed in Section 3.2, when software engineers collaborate and interact with other scientists and engineers, there could be challenges [9]. We experienced similar challenges as early as the requirements engineering phase. When software engineer members of the team started with use-case modeling, the notation and rationale was initially not clear for other team members. Furthermore, a much major issue started to occur, when domain experts started to explain the technical feature and specifications, e.g., about pumps and the fluid dynamics details. We observed that it was hard for the software engineers to follow and engage in such conversations. It was quite impractical for domain experts to learn SE concepts in detail, for software engineers to learn domain knowledge in depth. There was also a "terminology chasm" between experts with different backgrounds, which has also been reported in the grey literature [47]. With being open minded, what we found to be a good compromise was for each member to learn "just enough" about the topics s/he was not familiar with. Interestingly, this situation was similar to the "satisfice" metaphor [45], that we discussed above.

**Observation 3:** In scientific software projects, we experienced first-hand the widely-report experience that domain experts and software engineers often experience a kind of "terminology chasm". What we found to be a good and efficient way forward was for each member to learn "just enough" about the topics s/he was not familiar with.

### 5.3 SOFTWARE DESIGN AND DEVELOPMENT

Design and development activities were also conducted iteratively. Once an early prototype of design and development activities was ready, we improved it incrementally by involving all stakeholders in the project team.

As visualized in the SE process of Figure 2, we ensured progressing with the two intertwined sub-processes (domain research and SE) in parallel, and with close synchronization. While the domain research was the core activity for a while, the SE process started in parallel to it.

The software application was mainly designed to be a desktop Windows application. We also had planned the possibility to make the software available as a web-based tool in future (which is being explored as of this writing). Thus, we selected the Model-View-Controller (MVC) architecture pattern [48] to design the system. Development of OptimalPipeline was done in Visual C#.NET as it provided suitable features to develop the system and to enable integration with the four external systems, e.g., with Google Earth and the commercial SCADA systems. The source LOC of final stable version of OptimalPipeline was about 16 KLOC in C#.

We show in Figure 4 the simplified class diagram of the system, which is based on the MVC pattern [48]. Using MVC, it was possible to change the "view" layer from the desktop application to a web-based tool later on. The MVC pattern has also been used in development of other scientific software systems in the literature [49].

We observed that usage of the MVC pattern streamlined the implementation. It also supported design for testability [50, 51] by improving observability and controllability for the purpose of automated testing (discuss in Section 5.4). Furthermore, usage of the MVC pattern also supported maintainability [52], since we observed that maintaining the software (e.g., adding features to it) in later versions was not complex. A similar experience has also been reported by other researchers that the MVC pattern is a suitable choice of architecture, in general, for scientific software, e.g., [53-55].

In terms of development model, we chose to follow the rapid prototyping and Agile methods, as discussed in Section 5.1. Rapid prototyping allowed us to check both functional and non-functional requirements of the system in iterations among all tea members. For example, we were able to ensure high usability of the software in the development process.

Apart from the software in .Net, there was a major development effort related to the optimization model, which was developed in the domain research sub-process (Section 5.1). The optimization model was formulated and implemented using the commercial optimization engine LINGO/LINDO (www.lindo.com). Our choice of the optimization engine was made by an extensive review of available tools and also discussions with several optimization experts in and outside our team. For each of the case-study pipeline networks of our partners, a customized optimization script was developed using the optimization approaches that we reported in [8, 24]. The good news was that most of the optimization script was similar for different pipeline networks, and only a small part of it had to be customized for a given pipeline. For the case of one case-study pipeline of one of our partners, the optimization script was about 9 KLOC in the domain-specific scripting language of the LINGO tool. The script can be found in a thesis document [24].

While the GUI *OptimalPipeline* was a standard Windows application, the optimization script was to run by the optimization engine LINGO. To interconnect the two, we developed *OptimalPipeline* in a way to programmatically invoke the optimization script, pass the required parameters and get the output (optimization results).

In summary, while the MVC pattern was a good choice in our context, we realize that recommending use of MVC in designing every scientific software would not necessarily be a generalizable statement. We realized that the MVC pattern was a good choice since our system had a classical nature of data-application logic-graphical user interface (GUI), on which the MVC pattern works quite well. We could imagine other scientific software projects that are less GUI driven and hence will not require MVC but perhaps other pattern(s) [48] should be systematically chosen when designing them.

> **Observation 4:** When designing scientific software, developers should select a proper choice of architecture and architectural patterns [48]. In our context, by reviewing the literature and architectural guidelines, we selected the MVC pattern, which in our case, supported design for testability and design for maintainability.

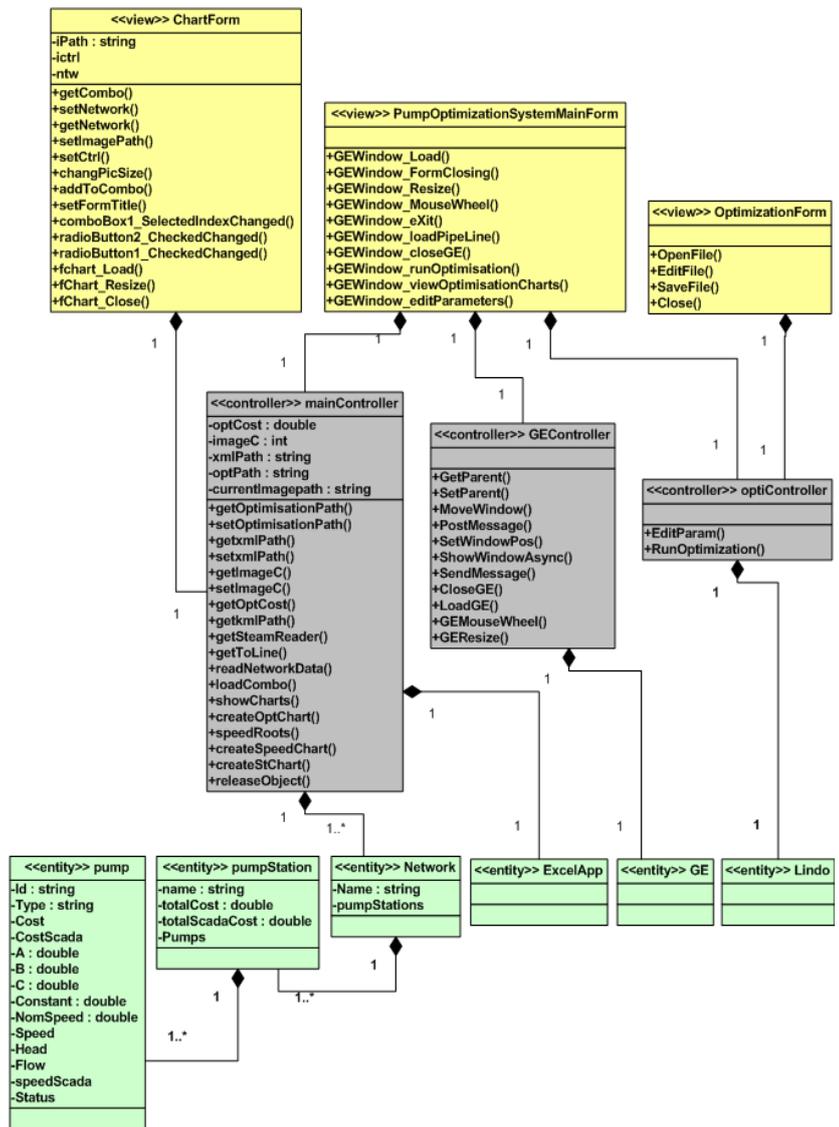

**Figure 4-System's class diagram (based on the MVC pattern)**

## 5.4 SOFTWARE TESTING

We used the test-driven development (TDD) practices, and this was a suitable practice since we were using Agile methods. We developed automated unit tests (in NUnit) for all methods of all classes in the *Model* and *Controller* levels of the MVC architecture. We also used a commercial record-playback GUI testing tool which was suitable for applications written in .Net (named Ranorex Studio). In choosing this GUI testing tool, we used heuristics to ensure that our choice was as systematic as possible [56]. The main factor was stability of the test tool and its applicability in our context (.Net environment).

For unit testing, test-case generation was done to ensure 100% code coverage. For GUI testing, we designed a GUI flow diagram the software under test (SUT) and ensured 100% path coverage on that diagram.

When developing test scripts, we ensured using best-practices in test-code development literature [57], e.g., we used the four-phase test pattern (setup, exercise, verify and teardown) in all our test methods. We observed no particular challenges in developing automated test scripts. The most challenge in testing was asserting the correctness of output of the optimization module, which we discuss next.

Testing of the subject software with respect to correctness of outputs of the optimization module was critical, since faults in this software could have major negative consequences. For instance, as discussed in Section 3.1, the *OptimalPipeline* software had to satisfy non-functional requirements with respect to minimum and maximum allowed pressure values in the oil pipeline. If the optimization module was to generate outputs in which the pressure of the liquid oil would go above the maximum or below the minimum threshold, catastrophic failures could occur, e.g., leakage and even explosion in the pipeline and pumping systems. Thus, the optimization module of the software had to fully satisfy such non-functional requirements. Criticality of effective testing for safety-critical scientific software has also been reported in other experience papers in the literature, e.g., in the context of oil reservoir simulation software [10], it was reported that: "*A more dangerous type of defect exists: faulty physics or functionality that can lead to erroneous simulation results*".

To plan and conduct testing activities on *OptimalPipeline*, we benefitted from the literature on testing scientific software [5]. We first identified the challenges and then laid down a suitable test strategy. We observed that there were inherent challenges concerning test-case design when testing the optimization module, due to the following example factors: (1) identifying critical input domain boundaries was often not trivial; (2) selecting a sufficient set of test cases was challenging due to the large number of input parameters and values accepted by the software; (3) since we were dealing with an optimization problem, we generally did not know the expected output for a given set of inputs (the test oracle problem).

We were able to apply some of classical black-box test-case design approaches (e.g., category-partitioning), e.g., we systematically changed minimum and maximum allowed pressure of each pipeline segment, and verified if *OptimalPipeline* would still provide valid outputs which would satisfy those ranges.

The next question in our testing efforts was; How do we verify if the outputs from the software (its optimization module) are really optimal (lowest pumping energy)? We found out that generation of test oracles for this system was also challenging due to various reasons, e.g.: (1) we developed the *OptimalPipeline* software to find the optimal operating schedule (optimization of pump-speed configurations) that were previously unknown; and (2) it was difficult (actually impossible most of the time) to determine the correct output of OptimalPipeline to verify the computational optimization model that involves complex calculations and simulations. Thus, as reported in the literature, a fundamental challenge in this context is: "*What if we could know that a program is buggy, even if we could not tell whether or not its observed output is correct?*" [58].

Let us recall from Section 3.1 that we formulated and solved the optimization problem using different optimization approaches, e.g.: (1) an Mixed-Integer Linear Programming (MILP) [24], and (2) a non-linear model based on Genetic Algorithms [8]. When testing the optimization model based on genetic algorithms [8], there were other inherent uncertainties due to randomness nature of genetic algorithms, since the program could return different outputs for the same values of inputs, in different executions.

> **Observation 5:** When a given scientific software has an optimization component, developers will expect challenges with respect to the test-oracle when testing the system.

To tackle the test-oracle challenge, we used a number of approaches. The first and simplest one was to compare the optimal pump-speed configurations, provided by OptimalPipeline, with the speed configurations already in use in the partner companies in the past, which were developed by expert engineers. It was obvious to expect that optimal results should be "better" (or at least ness worse), cost-wise, than speed configurations provided by experts.

Another popular approach for dealing with the test-oracle approach when finding out the expect outputs of a SUT is almost impossible (which was the case in our project), is metamorphic testing [5, 58]. Metamorphic testing operates by checking whether the SUT behaves according to a pre-defined set of properties known as metamorphic *relations*. A metamorphic relation (property) specifies the relationship in SUT's outputs/behavior between two different input values [5]. For example, in our project, one of the inputs to the software was the contractual delivery volume of oil products (in $m^3$). A metamorphic relation in this case was to increase the delivery volume across two test executions and expect that the operation cost of the pumps would go up in relative amounts (and would not decrease for example).

Let us note that while the concept of metamorphic testing is similar to "sensitivity analysis" in the optimization community, there are differences between the two and conducting both these types of validation techniques is often recommended in the literature, to ensure the quality of a software or model under analysis: "*While the primary objective of sensitivity analysis is to identify how sensitive a simulation system is to its different parameters, the objective of metamorphic testing is to look for anomalous behavior or outputs due to predefined input modifications or changed conditions, which could indicate a potential model defect*" [59].

We defined additional several metamorphic relations for metamorphic testing of our software (in addition to the above example) as we discuss next. Recall from Section 3.1 that our optimization software considered multi-tariff electricity supply as an input in the optimization operation. Thus we defined another metamorphic relation to verify if the electricity supply tariff was increased for a given time period (e.g., in the afternoon), the optimization software will provide "less" afternoon operation in the output schedule. Multiple additional metamorphic relations were defined corresponding to the physical and operational constraints of the pipeline (e.g., minimum and maximum allowed pressure of each pipeline segment, and hydraulic model). We systematically reduced those minimum and maximum values in different test runs, and verified if the output schedule produced by the optimization software still satisfied those constraints.

> **Observation 6:** When testing scientific software, for which determining the test-oracle is challenging or impossible, a suitable approach is to use metamorphic testing.

### 5.5 SOFTWARE MAINTENANCE

Since our development method was iterative, software maintenance was an integral part of the project. We conducted all four types of software maintenance:

- Adaptive: There were cases which we had to modify the system to cope with changes in the software environment, e.g., for different industry partners, the data format coming from SCADA systems were different and thus adaptive maintenance had to be done.
- Perfective: During different versions of the tool, we implemented new or changed features which concerned functional enhancements to the software, and often related to improvements in the underlying optimization model. For example, in several steps during the project, we expanded the output of the optimization model and algorithm by including the feedback from the domain experts, e.g., showing power-consumption charts for each pumping station (see Figure 5). In those stages, we conducted "perfective" maintenance on the OptimalPipeline software.
- Corrective: In several occasions, we diagnosed and fixed a number of defects, which were found either by our automated test suites or during manual testing / usage. As per our diagnosis, the sources (root causes) of defects could either be in the optimization model or the software itself. We carefully fixed those defects and re-executed the automated test suites to verify those features again. If there was a need to expand the automated test suites (add new test cases) to catch the latest defects and prevent them from appearing in future (regression testing), we did so.
- Preventive: Although we had used the MVC architecture pattern which supported software maintainability, we had to improve the architecture in a few occasions to increase software maintainability. For this, we mainly conducted refactoring.

> **Observation 7:** In a typical scientific software, often times, refinements and improvements in the domain research drive the need for software maintenance on the scientific software that is being developed, which in turn, serves (supports) the domain research.

### 5.6 COOPERATIVE AND HUMAN ASPECTS OF SOFTWARE ENGINEERING

As mentioned in an IEEE Software editorial [60]: "*… software isn't simply the product of appropriate technical solutions applied in appropriate technical ways. Software is developed by people, used by people, and supports interaction among people. As such, human characteristics, behavior, and cooperation are central to practical software development*". Since our project was a team of eight engineers from different disciplines, all the domain research (pipeline optimization) and SE activities were conducted by cooperative work. Thus, cooperative and human aspects of SE were visible in every phase of the project, e.g., as discussed in the requirements engineering phase above, there was also a "terminology chasm" among experts with different backgrounds, which has also been reported in by others in the literature [47]. Due to the terminology chasm among experts, we had to organize more than usual meetings in the early phases of the project so that team members could establish a common terminology set of communications and learning from each other's expertise, e.g., software engineers to learn some minima knowledge about pipelines and mechanical engineers learn a bit about SE. For example, software engineers in the team had to learn a bit about the domain concepts, e.g., pump parameters and characteristics such as pressure, capacity (flow-rate), efficiency and the hydraulic model linking all of them together. In turn, "other" engineers of the team (mechanical and power engineers) showed flexibility to learn about some of the SE concepts used in the process. We should note that all team members were passionate and ambitious about the project, thus there was inherent encouragements to support mutual learning.

Furthermore, there were also times when non-software-engineer members of the team did not see the need for systematic SE practices, e.g., using the MVC architecture. They would prefer just a simple (ad-hoc) design which would "just work", without in-depth thinking about design for testability and design for maintainability. We made a simple observation in testing, when non-software-engineer team members did not see the need for systematic test-case design. Previous studies have reported similar observations, e.g., a SLR [5] on testing scientific software reported that: "*The importance of unit testing is [often] not appreciated by scientists*".

In comparison to our other past projects in which all team members were software engineers [28], this particular project was clearly more challenging. However, by dedication to multi-disciplinary teamwork, open discussions and mutual learning, while the process was not easy at times, we were able to manage and overcome those differences, and achieve a successful outcome, an experience that has also been reported in other studies, e.g., [9, 43].

**Observation 8:** In a typical scientific software, when experts from different disciplines are involved, more efforts should be spent to ensure cooperation and knowledge sharing.

### 5.7 FINAL PRODUCT, COMMERCIALIZATION AND IMPACT

After all the efforts of team members, which were spent on domain research (optimization) and SE of the tool (*OptimalPipeline*), an initial version of the product was ready which we evaluated and improved together with industry partners.

Figure 5 shows a screenshot of *OptimalPipeline*, in which an optimization run has been completed and the system shows a chart comparing the cost of pumping based on optimal results versus historical SCADA data, for different pump stations.

Note that, due to confidentiality, this particular example is not a real pipeline network, but a hypothetical one. With the particular operating regime for pumps (as shown in Figure 5), the pipeline operator company is expected to achieve a saving of about $5K a day ($9K versus $4K) in pump energy consumptions. A video introduction to the software can be found in a YouTube video: www.youtube.com/watch?v=VF8cf38wNOQ

Since the product was carefully developed to meet the needs of industry partners in the project, a number of partners started using the product in their operations. We also conducted commercialization of the product in world-wide markets for a few years after the project.

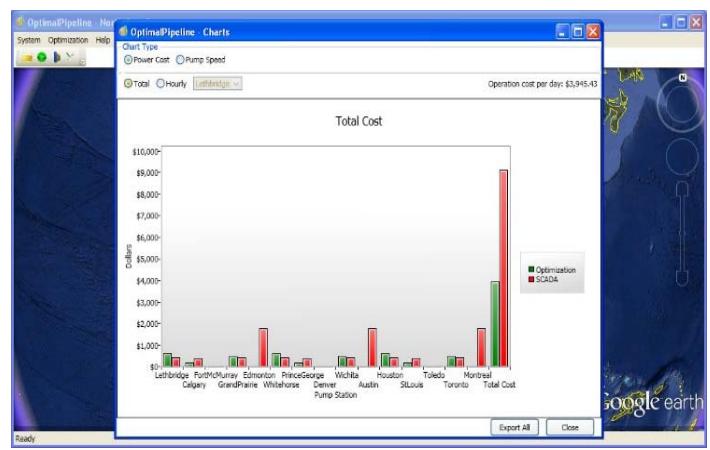

**Figure 5-A screenshot of the OptimalPipeline software**

### 6 SUMMARY AND DISCUSSION

We discuss the summary of our experience and implications, and then the limitations of our experience paper. In retrospective analysis of our experience in this 5-year project, we analyzed and reported our observations. We dealt with the challenges that we faced along the way, and our multi-disciplinary engineering approach was quite effective, for example, in the case of dealing with functional and non-functional requirements of the system under development.

We summarize below our lessons learnt and recommendations and provide take-away messages which would help other software engineers and scientists to better prepare for their scientific software development projects:

1. **Expect further challenges in projects having experts from different domains**: While having experts from different domains (mechanical, power and software engineers) was important for the success of our project, it made collaboration challenging at times. In comparison to our other past projects in which all team members were software engineers [28], this particular project was clearly more challenging. However, by dedication to multi-disciplinary teamwork, open discussions and mutual learning, we made the project successful.

2. **In scientific software, domain-specific requirements could often be more complex than use-case descriptions of the software**: We observed that most of the requirements and requirements engineering complexity was not in figuring out the software's use-cases, but instead lied in the domain-specific features and requirements of the system.
3. **Software engineering and domain research are closely interleaved**: Since the software was developed to enable automation of the domain research, the two activities were closely interleaved.
4. **Design for testability, maintainability and interoperability using suitable software architecture patterns**: Integration with other systems, such as SCADA, was made possible by careful design for interoperability using the Model-View-Controller (MVC) architecture pattern. Also, the MVC pattern made supported testability, maintainability. When designing scientific software, we observed that developers should select a proper choice of architecture and architectural patterns [48].
5. **Testing scientific software needs specific (non-traditional) testing approaches**: Testing and QA activities were also dealt with by including all team members and using suitable techniques such as metamorphic testing [5].
6. **Knowledge exchange and mutual learning**: At the end of the project, team members with different backgrounds mentioned their positive feelings about knowledge exchange and mutual learning. For example, of the mechanical and civil engineers in the team mentioned that they learned more about requirements engineering and testing in this project and they would be able to better communicate and collaborate with software engineers in future projects.
7. **Choosing, customizing and applying the "right" SE practices**: In some of the SE phases during the project, we opportunistically selected, customized and applied a number of SE practices which showed to provide benefits in the project.

We believe that the above lessons learnt and recommendations would be helpful for other software engineers and scientists in better preparing them for their projects related to development of scientific software. We should acknowledge that our above lessons learnt and recommendations are not exhaustive. Also, we should comment on the "perspective" from which the above lessons learned were synthesized: two of the authors are software engineers and one author is a domain expert (an expert in applied optimization). Thus, we have made sure to compile and synthesize the lessons learned from both perspective to minimize bias.

On the other hand, while he lessons learned were derived from one project, by reviewing the exiting body of experience in this area, e.g., [10-13], it is our belief that the lessons learnt and recommendations could be generalized to many scientific software development contexts. By adding to the exiting body of experience in this area, this experience is another step in utilizing SE best practices when developing scientific software.

We should note that we reported in this paper only a highlight of our SE practices and experiences in the subject project. More details can be found in a MSc thesis [61].

## 7 CONCLUSIONS AND FUTURE WORK

As our experience showed in this paper, development of scientific and engineering software is usually different and could be more challenging than the development of conventional enterprise software. Given the complex nature of the software (a sophisticated underlying optimization model), there were challenges in various software engineering aspects of the software system (e.g., requirements and testing). We reported our observations and experience in addressing those challenges during our technology-transfer project and we aimed to add to the existing body of experience and evidence in engineering of scientific and engineering software. We believe that our observations, experience and lessons learnt could be useful for other researchers and practitioners in engineering of scientific and engineering software.

The team members are continuing their discussions with scientists and they may start new projects such as the one reported in this paper. We have the vision to apply more of software engineering approaches in such projects and to increase the knowledge transfer from the software engineering discipline to community of scientists who develop software. Another future work of ours is to synthesize the vast number of experience reports in this area, adding our own experiences as reported in this paper, to be able to provide recommendations on usage of software engineering approaches in scientific and engineering software (when and what SE techniques to apply in a given scientific software project?).

## ACKNOWLEDGEMENTS

This work was supported by the Alberta Ingenuity New Faculty Award no. 200600673.

## REFERENCES

[1] A. Brett *et al.*, "Research Software Engineers: State of the Nation (UK) Report," *http://doi.org/10.5281/zenodo.495360*, 2017.

[2] J. Segal and C. Morris, "Developing Scientific Software," *IEEE Software,* vol. 25, no. 4, pp. 18-20, 2008.

[3] J. E. Hannay, C. MacLeod, J. Singer, H. P. Langtangen, D. Pfahl, and G. Wilson, "How do scientists develop and use scientific software?," in *Workshop on Software Engineering for Computational Science and Engineering*, 2009, pp. 1-8.

[4] R. Farhoodi, V. Garousi, D. Pfahl, and J. P. Sillito, "Development of scientific software: a systematic mapping," *International Journal of Software Engineering and Knowledge Engineering,* vol. 23, no. 04, pp. 463-506, 2013.

[5] U. Kanewala and J. M. Bieman, "Testing scientific software: A systematic literature review," *Information and Software Technology,* vol. 56, no. 10, pp. 1219-1232, 2014.

[6] University of Calgary, "Pipeline researchers get Alberta Ingenuity funding," *https://www.ucalgary.ca/news/uofcpublications/oncampus/biweekly/nov9-07/pipeline,* Last accessed: April 2019.

[7] E. Abbasi and V. Garousi, "An MILP-based Formulation for Minimizing Pumping Energy Costs of Oil Pipelines: Beneficial to both the Environment and Pipeline Companies," *Springer Journal on Energy Systems,* vol. 1, no. 4, pp. 393-416, 2010.

[8] E. Abbasi and V. Garousi, "Multi-Objective Optimization of both Pumping Energy and Maintenance Costs in Oil Pipeline Networks using Genetic Algorithms," in *Proceedings of the International Conference on Evolutionary Computation*, 2010, pp. 153-162.

[9] J. Segal, "When Software Engineers Met Research Scientists: A Case Study," *Empirical Software Engineering,* vol. 10, no. 1, pp. 517-536, 2005.

[10] M. v. Malkenhorst and L. Mollinger, "Going Underground," *IEEE Software,* vol. 29, no. 3, pp. 17-20, 2012.

[11] S. M. Easterbrook and T. C. Johns, "Engineering the Software for Understanding Climate Change," *Computing in Science & Engineering,* vol. 11, no. 6, pp. 65-74, 2009.

[12] J. Carver, "What We Have Learned about using Software Engineering Practices in Scientific Software," *A talk for the National Center for Supercomputing Applications (NCSA), https://www.youtube.com/watch?v=Hntri3kCJcI,* Last accessed: Mar. 2018.

[13] C. Haupt, M. Meinel, and T. Schlauch, "The software engineering initiative of DLR: overcome the obstacles and develop sustainable software," in *IEEE/ACM 13th International Workshop on Software Engineering for Science*, 2018, pp. 16-19.

[14] R. Farhoodi, V. Garousi, D. Pfahl, and J. P. Sillito, "Development of Scientific Software: A Systematic Mapping, Bibliometrics Study and a Paper Repository," *International Journal of Software Engineering and Knowledge Engineering,* vol. 23, no. 04, pp. 463-506, 2013.

[15] D. Heaton and J. C. Carver, "Claims about the use of software engineering practices in science: A systematic literature review," *Information and Software Technology,* vol. 67, pp. 207-219, 2015.

[16] M. T. Sletholt, J. Hannay, D. Pfahl, H. C. Benestad, and H. P. Langtangen, "A literature review of agile practices and their effects in scientific software development," in *Proceedings of the International Workshop on Software Engineering for Computational Science and Engineering*, 2011, pp. 1-9.

[17] T. Dyba, B. A. Kitchenham, and M. Jorgensen, "Evidence-based software engineering for practitioners," *IEEE software,* vol. 22, no. 1, pp. 58-65, 2005.

[18] K. Petersen and C. Wohlin, "Context in industrial software engineering research," in *International Symposium on Empirical Software Engineering and Measurement*, 2009, pp. 401-404.

[19] T. Dybå, D. I. Sjøberg, and D. S. Cruzes, "What works for whom, where, when, and why?: on the role of context in empirical software engineering," in *Proceedings of the ACM-IEEE international symposium on Empirical software engineering and measurement*, 2012, pp. 19-28.

[20] Central Intelligence Agency (CIA), "World Factbook 2014," *https://www.cia.gov/library/publications/download/download-2014/,* Last accessed: April 2019.

[21] A. Bahadori, *Oil and gas pipelines and piping systems: Design, construction, management, and inspection*. Gulf Professional Publishing, 2016.

[22] D. Trung, "Modern SCADA systems for oil pipelines," in *Annual Conference pf tje Petroleum and Chemical Industry Applications Society*, 1995, pp. 299-305.

[23] S. Pezeshk and O. Helweg, "Adaptive search optimization in reducing pump operating costs," *Journal of Water Resources Planning and Management,* vol. 122, no. 1, pp. 57-63, 1996.

[24] E. Abbasi, "Development and industrial application of an MILP-based optimization algorithm for minimizing pumping cost and carbon footprint of oil pipelines," *MSc thesis, University of Calgary,* 2010.

[25] T. Gorschek, P. Garre, S. Larsson, and C. Wohlin, "A model for technology transfer in practice," *IEEE software,* vol. 23, no. 6, pp. 88-95, 2006.

[26] D. E. Avison, F. Lau, M. D. Myers, and P. A. Nielsen, "Action research," *Communications of the ACM,* vol. 42, no. 1, pp. 94-97, 1999.

[27] R. M. Emerson, R. I. Fretz, and L. L. Shaw, "Participant observation and fieldnotes," *Handbook of ethnography,* pp. 352-368, 2001.

[28] V. Garousi, M. M. Eskandar, and K. Herkiloğlu, "Industry-academia collaborations in software testing: experience and success stories from Canada and Turkey," *Software Quality Journal,* vol. 25, no. 4, pp. 1091–1143, 2017.

[29] P. Runeson and M. Höst, "Guidelines for conducting and reporting case study research in software engineering," *Empirical Software Engineering,* vol. 14, no. 2, pp. 131-164, 2009.

[30] D. I. Sjoberg, T. Dyba, and M. Jorgensen, "The future of empirical methods in software engineering research," in

[30] *Conference on Future of Software Engineering* 2007, pp. 358-378.

[31] R. Davison, M. G. Martinsons, and N. Kock, "Principles of canonical action research," *Information systems journal,* vol. 14, no. 1, pp. 65-86, 2004.

[32] P. S. M. d. Santos and G. H. Travassos, "Action-research use in software engineering: An initial survey," in *Proceedings of the International Symposium on Empirical Software Engineering and Measurement*, 2009, pp. 414-417.

[33] C. Argyris and D. A. Schon, *Theory in practice: Increasing professional effectiveness*. Jossey-Bass, 1974.

[34] V. Garousi *et al.*, "Experience in automated testing of simulation software in the aviation industry," *IEEE Software,* July/August 2019.

[35] V. Garousi, "Experience in Developing a Robot Control Software," *Canadian Journal on Computer and Information Science,* vol. 4, no. 1, pp. 3-13, 2011.

[36] V. Garousi, D. C. Shepherd, and K. Herkiloğlu, "Successful engagement of practitioners and software engineering researchers: Evidence from 26 international industry-academia collaborative projects," *IEEE Software, In press,* 2019.

[37] A. J. Ko, "A three-year participant observation of software startup software evolution," in *Proceedings of IEEE/ACM International Conference on Software Engineering: Software Engineering in Practice Track*, 2017, pp. 3-12.

[38] S. Patil, A. Kobsa, A. John, and D. Seligmann, "Methodological reflections on a field study of a globally distributed software project," *Information and Software Technology,* vol. 53, no. 9, pp. 969-980, 2011.

[39] S. Kumar and C. Wallace, "Among the agilists: Participant observation in a rapidly evolving workplace," in *Proceedings of International Workshop on Cooperative and Human Aspects of Software Engineering*, 2016, pp. 52-55.

[40] C. B. Seaman, "Qualitative methods in empirical studies of software engineering," *IEEE Transactions on software engineering,* vol. 25, no. 4, pp. 557-572, 1999.

[41] D. E. Knuth, "The errors of TeX," *Software: Practice and Experience,* vol. 19, no. 7, pp. 607-685, 1989.

[42] D. W. Kane, M. M. Hohman, E. G. Cerami, M. W. Mccormick, K. F. Kuhlmman, and J. A. Byrd, "Agile methods in biomedical software development: a multi-site experience report," *Bioinformatics,* vol. 7, no. 1, pp. 273-285, 2006.

[43] D. Kane, "Introducing Agile Development into Bioinformatics: An Experience Report," in *Proceedings of the Conference on Agile Development*, 2003, pp. 132-140.

[44] A. Davis, *Just enough requirements management: where software development meets marketing*. Addison-Wesley, 2013.

[45] A. Tang and H. van Vliet, "Software designers satisfice," in *European Conference on Software Architecture*, 2015, pp. 105-120.

[46] L. J. Ball, L. Maskill, and T. C. Ormerod, "Satisficing in engineering design: Causes, consequences and implications for design support," *Automation in construction,* vol. 7, no. 2-3, pp. 213-227, 1998.

[47] S. J. Runge, "Bridging the IT and Business Terminology Chasm," *https://itpsb.blogspot.com/2012/01/bridging-it-and-business-terminology.html*, Last accessed: April 2019.

[48] L. Bass, P. Clements, and R. Kazman, *Software architecture in practice*. Addison-Wesley Professional, 2003.

[49] M. D. Pierro, "web2py for Scientific Applications," *Computing in Science & Engineering,* vol. 13, no. 2, pp. 64-69, 2011.

[50] D. Esposito, "Chapter 9. Testing and testability in ASP.NET MVC," in *Programming Microsoft ASP.NET MVC*: Microsoft Press, 2014.

[51] V. Garousi, M. Felderer, and F. N. Kılıçaslan, "A survey on software testability," *Information and Software Technology,* vol. 108, pp. 35-64, 2019.

[52] A. Meadows, "Chapter 6. Writing Maintainable Code," in *Chapter 6. Writing Maintainable Code*: Packt Publishing, 2013.

[53] A. O. Fabiyi, "A methodology for developing scientific software applications in science gateways: towards the easy accessibility and availability of scientific applications," Brunel University London, PhD thesis, 2017.

[54] R. A. Lotufo, R. C. Machado, A. Körbes, and R. G. Ramos, "Adessowiki on-line collaborative scientific programming platform," in *Proceedings of international symposium on Wikis and open collaboration*, 2009, p. 10.

[55] N. V. Iannotti, "Improving reuse in software development for the life sciences," *Purdue University, PhD thesis,* 2013.

[56] P. Raulamo, M. V. Mäntylä, and V. Garousi, "Choosing the right test automation tool: a grey literature review," in *International Conference on Evaluation and Assessment in Software Engineering*, Karlskrona, Sweden, 2017, pp. 21-30.

[57] V. Garousi, Y. Amannejad, and A. Betin-Can, "Software test-code engineering: a systematic mapping," *Journal of Information and Software Technology,* vol. 58, pp. 123–147, 2015.

[58] S. Segura, D. Towey, Z. Q. Zhou, and T. Y. Chen, "Metamorphic Testing: Testing the Untestable," *IEEE Software,* 2018.

[59] M. Raunak and M. Olsen, "Simulation validation using metamorphic testing," in *Proceedings of the Conference on Summer Computer Simulation*, 2015, pp. 1-6.

[60] C. R. de Souza, H. Sharp, J. Singer, L.-T. Cheng, and G. Venolia, "Guest Editors' Introduction: Cooperative and Human Aspects of Software Engineering," *IEEE software,* vol. 26, no. 6, pp. 17-19, 2009.

[61] R. Farhoodi, "A systematic literature review of software engineering for scientific and engineering software and an industrial oil pipeline software case-study," *MSc thesis, University of Calgary,* 2010.